\documentclass[pre,twocolumn]{revtex4}
\usepackage{psfig}
\usepackage{epsfig}
\usepackage{amsfonts}
\usepackage{amsmath}
\usepackage{amssymb}
\usepackage{graphicx}

\input{tcilatex}

\begin{document}

\title{Geographical effects on cascading breakdowns of scale-free networks}
\author{Liang Huang$^{2}$, Lei Yang$^{2,3}$ and Kongqing Yang$^{1,2}$}
\affiliation{$^{1}$Institute of Applied Physics, Jimei University, Xiamen 361021, China}
\affiliation{$^{2}$Department of Physics, Lanzhou University, Lanzhou 730000, China}
\affiliation{$^{3}$Center for Nonlinear studies, and The Beijing-Hong Kong-Singapore
Joint Center for Nonlinear and Complex Systems (HK Branch), Hong Kong
Baptist University, Hong Kong, China}

\begin{abstract}
Cascading breakdowns of real networks are severe accidents in recent years,
such as the blackouts of the power transportation networks in North America.
In this paper, we study the effects of geographical structure on the
cascading phenomena of load-carried scale-free networks, find that more
geographically constrained networks tend to have larger cascading
breakdowns. Explanations by the effects of circles and large betweenness of
small degree nodes are provided.
\end{abstract}

\date{\today}
\maketitle

Recently dynamical processes on networks has been highly concerned and
widely investigated \cite{ws,watts0,strogatz}. Among many of the dynamical
features of networks, robustness attracts much attention \cite%
{bai,cohenr,coheni,newmanr}, much of which focus on scale-free (SF)
networks, i.e., the degrees of nodes satisfy a power law distribution: $%
P(k)\sim k^{-\lambda }$, for their ubiquity in real systems \cite{CN-review}%
. The heterogeneity of the degrees often makes the scale-free networks
sensitive to intentional attack \cite{coheni,newmanr}, while it is
resilience to random breakdowns \cite{cohenr,newmanr}, and also resilience
under avalanche phenomena by the role of the hubs that sustain large amounts
of grain, playing the role of reservoirs \cite{sdp}. Furthermore, for
cascading failures, the load-carried SF network is fragile even when one
attacks only one node or very few nodes with the largest degrees \cite%
{cascade}.

Since many real networks exist in two or three dimensional physical spaces,
it is helpful to study the geographical complex networks and it has
attracted much attention recently \cite{lesf,snsf,GCN-other,GCN-yang,geoperc}%
. It has been shown that geographical structure has great influence on
percolation thresholds \cite{geoperc}. Since many real systems bear
cascading failures, such as power grid networks, traffic lines, Internet,
etc., and also lay on the two dimensional global surface, the influence of
geographical structures on cascading breakdowns is of highly importance and
up to now is rarely studied.

In this paper, we study the effects of geographical structure on the
cascading phenomena of load-carried scale-free networks, in which each node
carries a certain type of load, such as power, traffic, etc., and if the
node is broken down, its load will be redistributed to its neighbors. We
investigate the Bak-Tang-Wiesenfeld (BTW) sandpile model \cite{sdp,btw} as a
prototypical model on a weighted lattice embedded SF (WLESF) network \cite%
{GCN-yang}; and further study the betweenness distribution. Both validate
that the more spatially loosely connected network is more robust under
cascading failures, i.e., they have less huge avalanche events. The network
is generated as follows \cite{GCN-yang}. It begins with an $L\times L$
lattice, with periodical boundary conditions, and for each node assigned a
degree $k$ drawn from the prescribed SF degree distribution $P(k)\sim
k^{-\lambda }$, $k\geqslant m$. Then a node $i$ is picked out randomly,
according to a Gaussian weight function $f_{i}(r)=De^{-\left( \frac{r}{A%
\sqrt{k_{i}}}\right) ^{2}}$, it selects other nodes and establishes
connections until its degree quota $k_{i}$ is filled or until it has tried
many enough times, avoiding duplicate connections. The process is carried
out for all the nodes in the lattice. The clustering parameter $A$ controls
the spatial denseness of the connections. For large $A$ limits, e.g. $A\sqrt{%
m}\gg L$, the weight function will be trivial, and the network becomes a SF
random (SFR) network, i.e., random otherwise than SF degree distribution 
\cite{grn}. To compare, we also investigated lattice embedded SF (LESF)
networks with nearest neighbor connections \cite{lesf}. Here, we assume that
the time scales governing the dynamics are much smaller than that
characterizing the network evolvement, thus the static geographical network
models are suitable for discussing the problems.

The rules we adopted for sandpile dynamics are as follows: (i) At each time
step, a grain is added at a randomly chosen node $i$. (ii) If the height at
the node $i$ reaches or exceeds a prescribed threshold $z_{i}=k_{i}$, the
degree of the node $i$, then it becomes unstable and all the grains at the
node topple to its adjacent nodes: $h_{i}=h_{i}-k_{i}$; and for each $i$'s
neighbor $j$: $h_{j}=h_{j}+1$; during the transfer, there is a small
fraction $f$ of grains being lost, which plays the role of sinks without
which the system becomes overloaded in the end. (iii) If this toppling
causes any of the adjacent nodes to be unstable, subsequent topplings follow
on those nodes in parallel until there is no unstable node left, forming an
avalanche event. (iv) Repeat (i) --(iii).

The main feature of the BTW sandpile model on Euclidean space is the
emergence of a power law with an exponential cutoff in the avalanche size
distribution, $p(s)\sim s^{-\tau ^{\prime }}e^{-s/s_{c}}$, where $s$ is the
avalanche size, i.e., the number of toppling nodes in an avalanche event,
and $s_{c}$ is its characteristic size. In our studies, nodes toppled more
than once in an avalanche event is seldom \cite{sdp}, unless for the very
large avalanches, which have already exceeded the exponential cutoffs. Thus
we study the avalanche area, which is the number of distinct nodes that
toppled in an avalanche event, instead of avalanche size. The avalanche area
distribution follows the same form as that of avalanche sizes%
\begin{equation}
p(a)\sim a^{-\tau }e^{-a/a_{c}},  \label{ps}
\end{equation}%
where $a$ is the avalanche area, and $a_{c}$ its characteristic size. A
typical example is shown in Fig. \ref{sa}.

\FRAME{ftbpFU}{7.5278cm}{6.1066cm}{0pt}{\Qcb{Number of avalanches with size $%
j$ or area $j$, for LESF networks out of $10^{6}$ avalanche events on one
network configuration. $m=4$, $N=10^{5}$.}}{\Qlb{sa}}{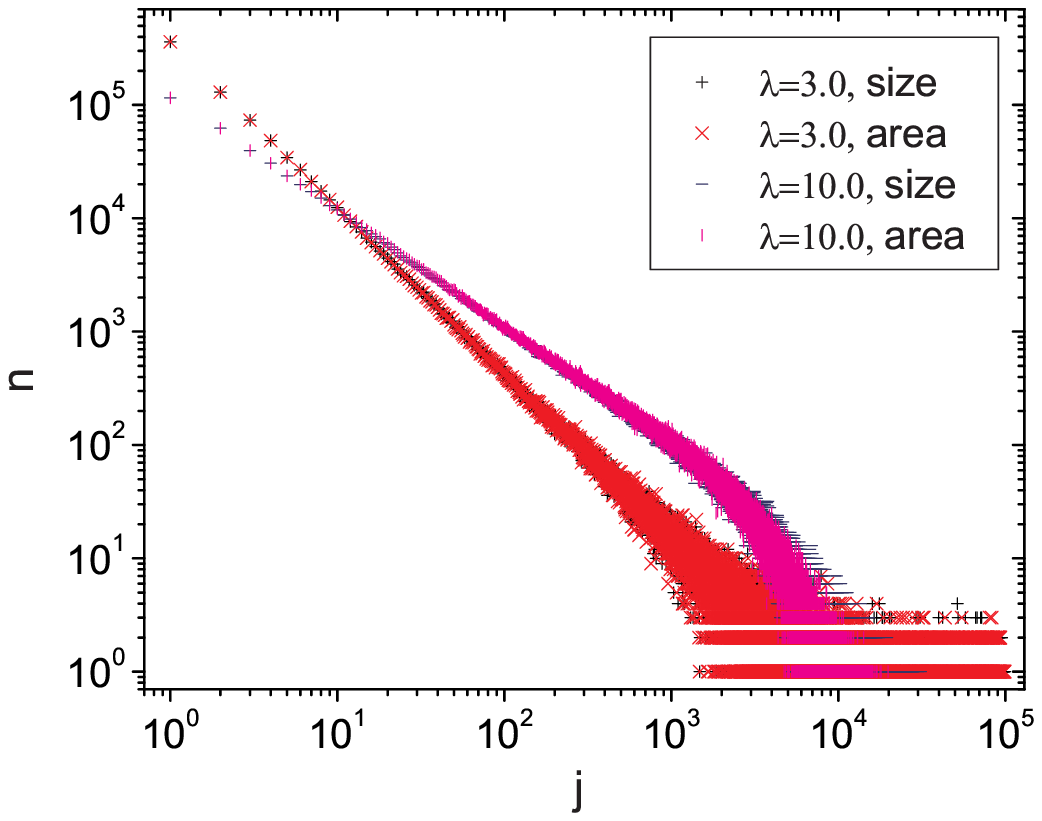}{\special%
{language "Scientific Word";type "GRAPHIC";maintain-aspect-ratio
TRUE;display "ICON";valid_file "F";width 7.5278cm;height 6.1066cm;depth
0pt;original-width 4.8049in;original-height 3.8977in;cropleft "0";croptop
"1";cropright "1";cropbottom "0";filename 'Fig1.EPS';file-properties
"XNPEU";}}

For BTW sandpile model on SFR networks, K. S. Goh \textit{et al.} \cite{sdp}
have shown that the avalanche area exponent $\tau $ increases as $\lambda $
decreases, caused by the increasing number of hubs playing the role of
reservoirs. Here, we will demonstrate that for the densely connected
scale-free geographical networks, the reservoir effect is weakened, and the
network has a smaller $\tau $.

Figure \ref{chnn} represents the avalanche area distribution for different $%
\lambda $ of LESF networks and WLESF networks with $A=1$. It shows that as $%
\lambda $ decreases, the curve of avalanche area distribution is steeper,
corresponds to larger $\tau $. These are the same as the results in Ref. 
\cite{sdp}. The avalanche area exponent $\tau $ for these data are fitted by
formula \ref{ps}, and is presented in Fig. \ref{yA1tau}, together with that
of SFR networks for comparison. The data for SFR networks we obtained is
consistent with that of \cite{sdp}. For large $\lambda $ large $N$ limits,
the SFR network tends to ER random graphs, for which $\tau \simeq 1.5$ \cite%
{sdp,bona}; while LESF network tends to a super lattice, with each node has $%
m$ neighbors; since in our studies $m=4$, the network limits to a normal $2D$
lattice, which has a value of $1.01(2)$ for $\tau $, consistent with the
previous results \cite{btw,OFC}.

\FRAME{ftbpFUw}{9.1006cm}{5.9748cm}{0pt}{\Qcb{Avalanche area distribution
for LESF (left panel) and WLESF $A=1$ (right panel) networks. For both
panels, from up to down $\protect\lambda =10.0$, $5.0$, $4.0$, $3.5$, $3.0$, 
$2.8$, $2.6$, $2.4$. The loosing probability is $f=0.001$, and $m=4$, $%
N=10^{5}$. $10$ network realizations are carried out and for each $10^{6}$
avalanche events are recorded for statistics.}}{\Qlb{chnn}}{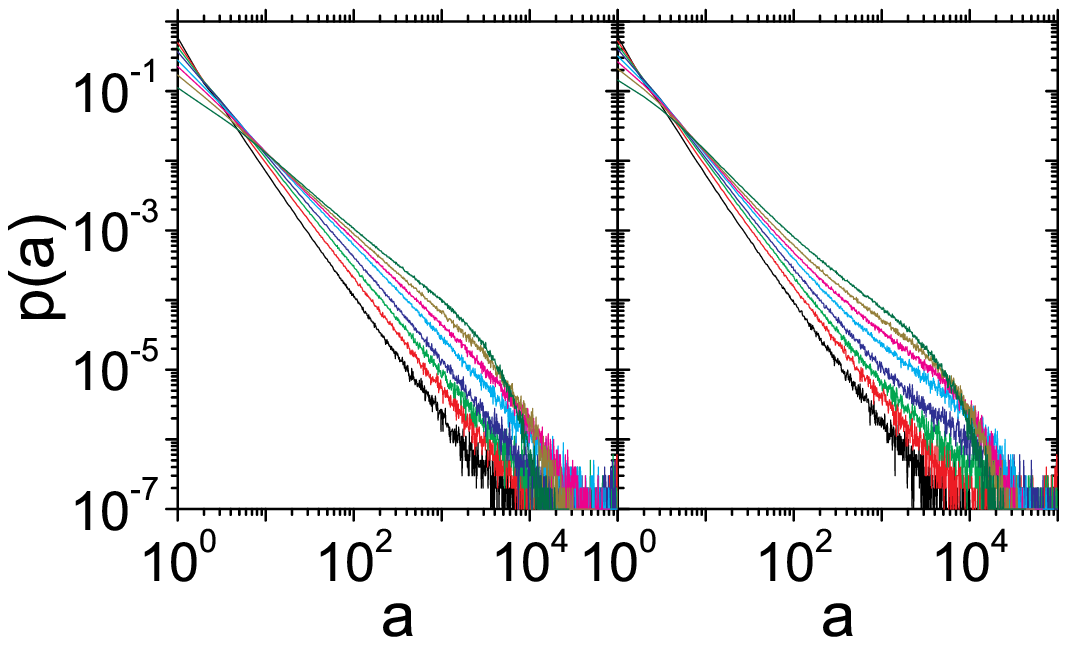}{%
\special{language "Scientific Word";type "GRAPHIC";maintain-aspect-ratio
TRUE;display "USEDEF";valid_file "F";width 9.1006cm;height 5.9748cm;depth
0pt;original-width 5.0678in;original-height 3.8977in;cropleft "0";croptop
"1";cropright "1";cropbottom "0";filename 'Fig2.EPS';file-properties
"XNPEU";}}

\FRAME{ftbpFU}{7.5981cm}{5.7178cm}{0pt}{\Qcb{Avalanche area exponent $%
\protect\tau $ vs the SF degree exponent $\protect\lambda $. The data are
fitted by formula \protect\ref{ps}, from the data presented in Fig.  \protect
\ref{chnn} and that of SFR networks. The network parameters and the
statistics for SFR network are the same as that in Fig. \protect\ref{chnn}.}%
}{\Qlb{yA1tau}}{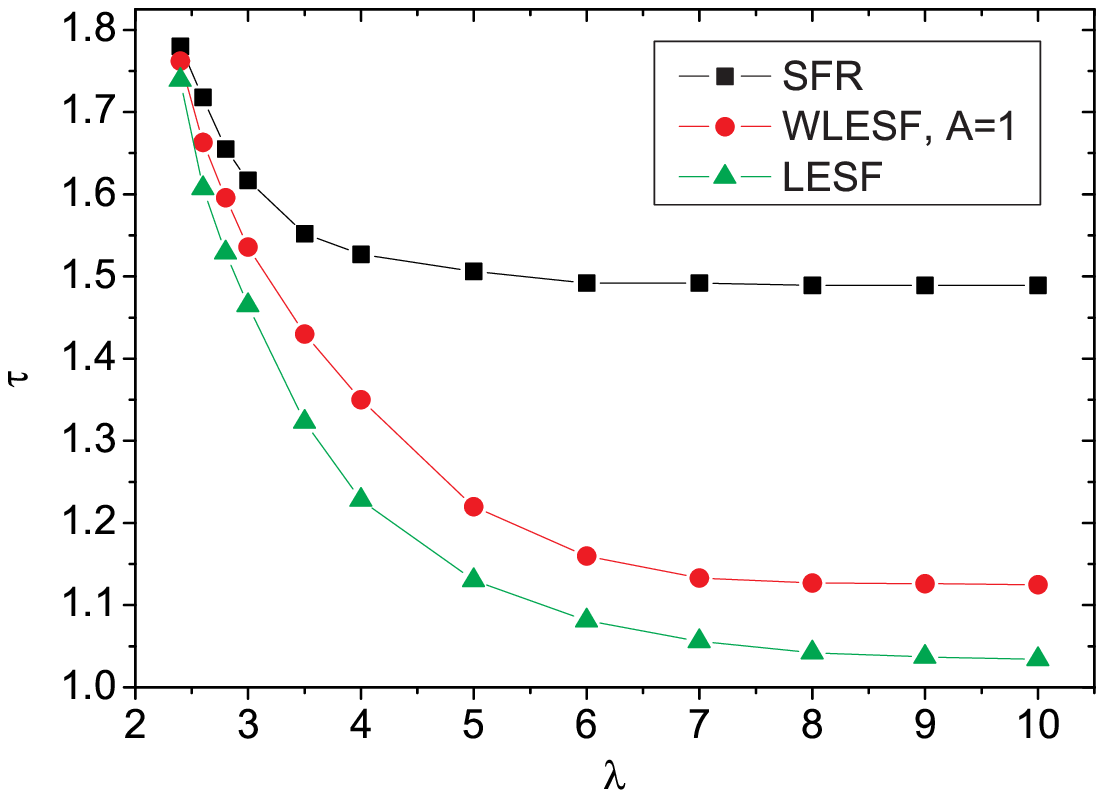}{\special{language "Scientific Word";type
"GRAPHIC";maintain-aspect-ratio TRUE;display "USEDEF";valid_file "F";width
7.5981cm;height 5.7178cm;depth 0pt;original-width 5.0825in;original-height
3.8112in;cropleft "0";croptop "1";cropright "1";cropbottom "0";filename
'Fig3.EPS';file-properties "XNPEU";}}

The avalanche area exponent for different $A$ of WLESF network is shown in
Fig. \ref{yan}. As $A$ goes larger, avalanche area exponent $\tau $
increases, the curves of avalanche area distribution become sharper in the
double-log plot (see inset of Fig. \ref{yan}), which corresponds to fewer
large avalanche events. This transition in $\tau $ illuminates that when the
network is geographically more loosely connected, it will be harder for
large cascading events to occur.

\FRAME{ftbpFU}{7.5981cm}{5.8189cm}{0pt}{\Qcb{Avalanche area exponent $%
\protect\tau $ vs the clusterness parameter $A$, for $\protect\lambda =3.0$
(squares), $5.0$ (circles) and $10.0$ (triangles). The data are fitted by
formula \protect\ref{ps}. Inset: Avalanche area distribution for $\protect%
\lambda =3.0$, from top to bottom are LESF, WLESF $A=1$, $A=2$, and SFR
networks. Dynamical and network parameters are the same as that in Fig. 
\protect\ref{chnn}.}}{\Qlb{yan}}{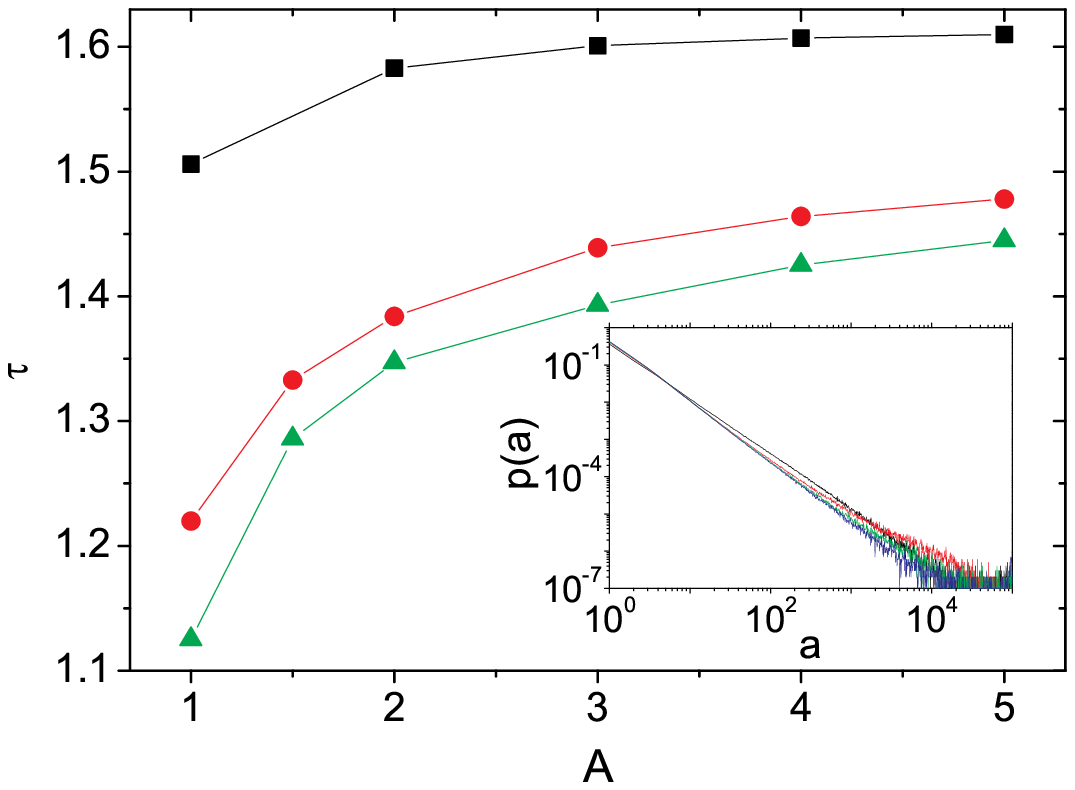}{\special{language "Scientific
Word";type "GRAPHIC";maintain-aspect-ratio TRUE;display "USEDEF";valid_file
"F";width 7.5981cm;height 5.8189cm;depth 0pt;original-width
5.0125in;original-height 3.8277in;cropleft "0";croptop "1";cropright
"1";cropbottom "0";filename 'Fig4.EPS';file-properties "XNPEU";}}

The range of an edge is the length of the shortest paths between the nodes
it connected in the absence of itself \cite{watts0,range}. If an edge's
range is $l$, it will probably belong to an $l+1$ circle. Thus the
distribution of range in a network sketches the distribution of circles. The
inset of Fig. \ref{sdpf} shows that when the spatial constrains is slighter,
as $A$ goes larger, the range distribution drifts to larger ranges. It means
that spatially loosely connected networks have fewer small order circles but
more higher order circles. If there are many small order circles, the
toppling grains are easier to meet, and the nodes with much less grains,
i.e., fewer than $z-1$, especially those with $z-2$ or $z-3$ grains, could
also reach the toppling threshold $z$ and topple. Larger order circles
contribute less to this effect. The main frame of Fig. \ref{sdpf} shows the
fraction of nodes toppled in avalanches that have precisely $z-1$ grains. As
the network is less geographically constrained and has fewer small order
circles, the fraction of toppling nodes with $z-1$ grains increases,
justifies our reasoning. This effect contributes to the large avalanche
events of the densely connected networks, and explains the decrease of
avalanche area exponent $\tau $ as the network is more geographically
constrained.

\FRAME{ftbpFU}{7.5981cm}{6.1857cm}{0pt}{\Qcb{Fraction of nodes that toppled
after receiving only one grain in an avalanche event vs avalanche area. From
bottom to top is LESF (squares), WLESF $A=1$ (circles), $A=2$ (up
triangles), $A=3$ (down triangles), $A=5$ (diamonds), and SFR network (left
triangles). Each has $10^{6}$ avalanche records on one network for
statistics. $\protect\lambda =3$, $m=4$,\ $N=10^{5}$. Inset: Range
distribution of the same networks; same symbols represent same networks as
that in the main frame.}}{\Qlb{sdpf}}{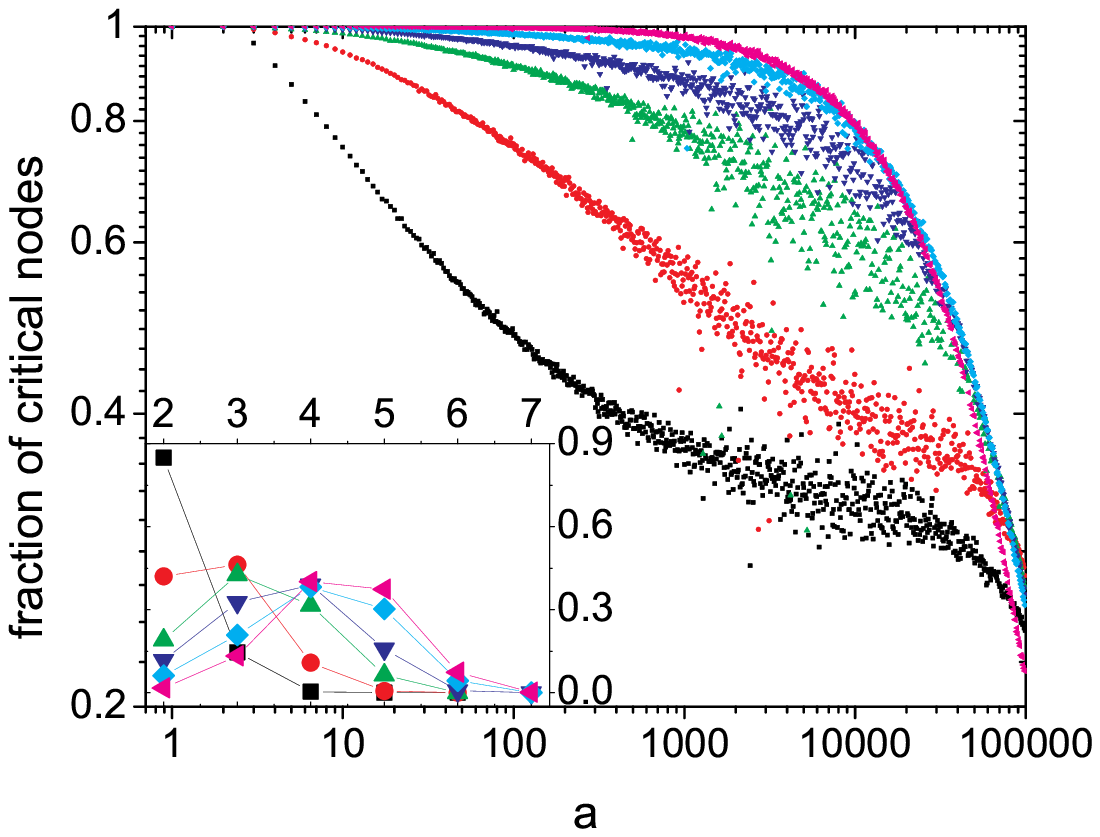}{\special{language
"Scientific Word";type "GRAPHIC";maintain-aspect-ratio TRUE;display
"USEDEF";valid_file "F";width 7.5981cm;height 6.1857cm;depth
0pt;original-width 5.1378in;original-height 3.9375in;cropleft "0";croptop
"1";cropright "1";cropbottom "0";filename 'Fig5.EPS';file-properties
"XNPEU";}}

In the following section, we studied the betweenness distribution of these
geographical networks. The betweenness, or betweenness centrality, of node $%
i $ is defined as the total number of shortest paths between pairs of nodes
that pass through $i$ \cite{btnc}. If a pair of nodes has two shortest
paths, the nodes along those paths are given a betweenness of $1/2$ each.
The betweenness distribution for SF networks is reported to follow a power
law $P_{B}\sim b^{-\delta }$, and for $2<\lambda \leqslant 3$, the exponent
is $\delta \approx 2.2(1)$ \cite{sdp}. We find that the betweenness
distribution of LESF network decays much slower than that of SFR networks,
as Fig. \ref{btnd} demonstrates for a particular case. The distributions for
WLESF networks lay between them, but do not appear in the graph for
clearness. The same holds for other $\lambda $ and $m$ values. Thus there
are more large betweenness nodes in LESF networks than in SFR networks. To
comprehend this, we plot the betweenness vs node's degree in Fig. \ref{btndt}%
. For LESF networks the betweenness of the same degree is distributed much
more diffusively, and on average are larger. It could be seen that even
nodes with small degree $k$ could have unusually large betweenness.

\FRAME{ftbpFU}{7.5981cm}{6.1989cm}{0pt}{\Qcb{Betweenness distribution of the
networks. $\protect\lambda =3.0$, $m=2$, and network size $N=10^{4}$, each
has been averaged over $100$ configurations.}}{\Qlb{btnd}}{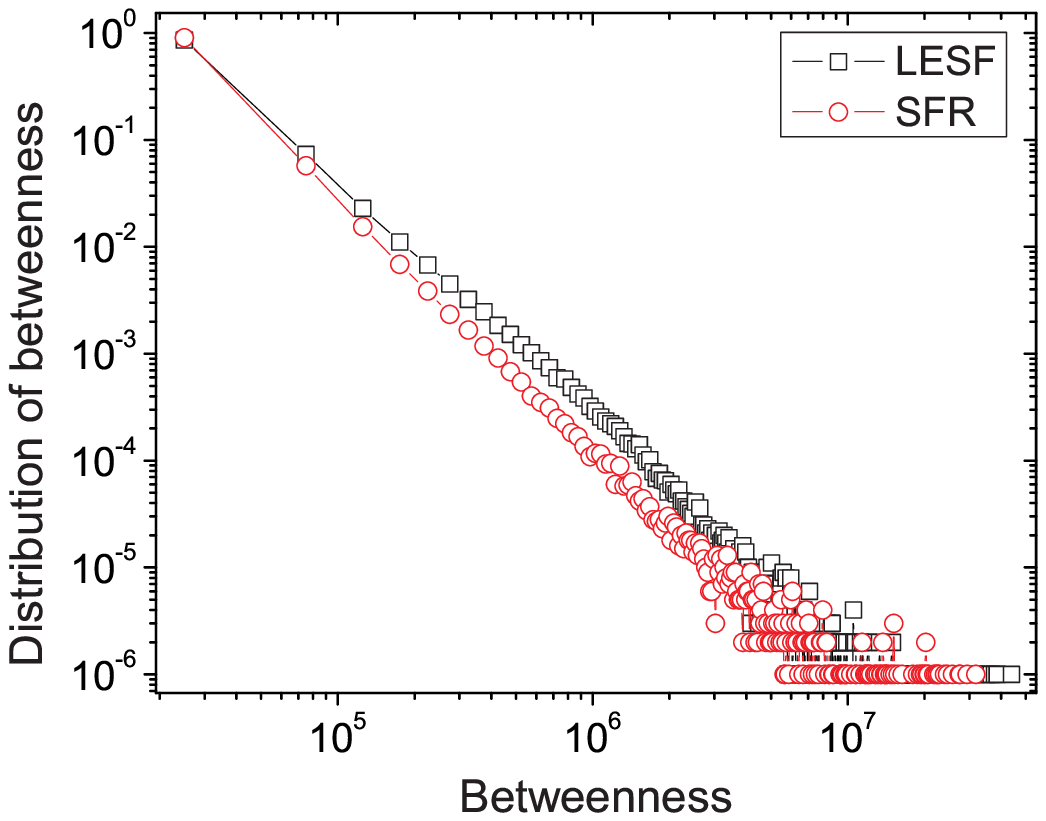}{\special%
{language "Scientific Word";type "GRAPHIC";maintain-aspect-ratio
TRUE;display "USEDEF";valid_file "F";width 7.5981cm;height 6.1989cm;depth
0pt;original-width 4.8049in;original-height 3.9098in;cropleft "0";croptop
"1";cropright "1";cropbottom "0";filename 'Fig6.EPS';file-properties
"XNPEU";}}

\FRAME{ftbpFU}{7.5278cm}{6.1418cm}{0pt}{\Qcb{Betweenness $b$ vs \ degree $k$
of nodes. Data are the same as that in Fig. \protect\ref{btnd}.}}{\Qlb{btndt}%
}{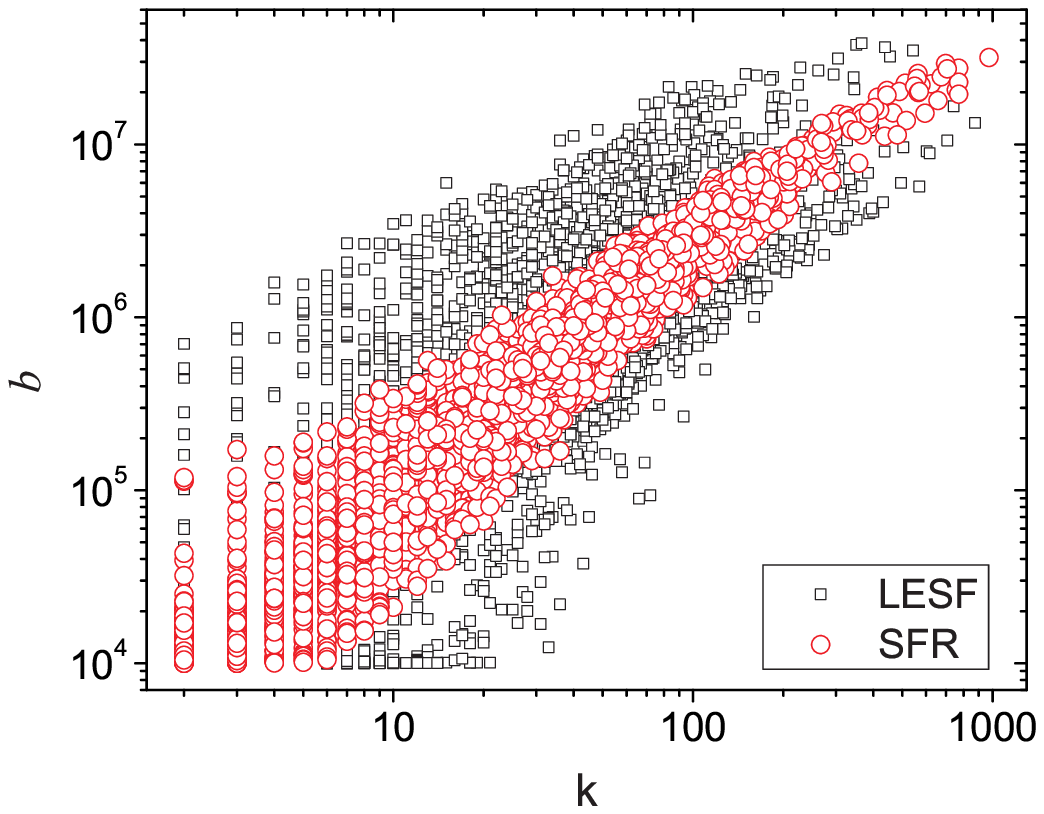}{\special{language "Scientific Word";type
"GRAPHIC";maintain-aspect-ratio TRUE;display "ICON";valid_file "F";width
7.5278cm;height 6.1418cm;depth 0pt;original-width 5.1378in;original-height
3.9375in;cropleft "0";croptop "1";cropright "1";cropbottom "0";filename
'Fig7.EPS';file-properties "XNPEU";}}

When an avalanche occurs, the front of toppling nodes spread along
geodesics, i.e., along the shortest paths between nodes. Since the
betweenness of a node is the number of shortest paths passing through it,
larger betweenness means that it will have higher possibility to receive
grains in avalanching processes. In the above sandpile model, the toppling
threshold is the node's degree, thus the node that has large betweenness but
small degree will be easier to topple. As Fig. \ref{btndt} shows, LESF
network have more such nodes than SFR networks, and the situation changes
continuously for WLESF network with increasing $A$. This could also account
for the decreasing avalanche area component $\tau $ as the network is more
geographically constrained.

In conclusion, by studying avalanching processes on geographical SF
networks, we find that besides the reservoir effects of the hubs in SF
networks, geography has great influences on the critical exponents of these
systems. The decreasing avalanche area component $\tau $ for the more
geographically constrained network hints high risks for such network to
breakdown through cascading failures, since they have a much higher
possibility to experience huge avalanche events, due to the denser
connections and huge number of smaller order circles and larger betweenness
of small degree nodes. Since many real networks that carried some kinds of
loads, i.e., power, traffic, data packets, etc., are imbedded in the $2D$
global surface and highly clustered, our results indicate that they will
suffer more severe risks under node failures.

The work is supported by China National Natural Sciences Foundation with
grant 49894190 of major project and Chinese Academy of Science with grant
KZCX1-sw-18 of major project of knowledge innovation engineering. L. Yang
thanks the support of the Hong Kong Research Grants Council (RGC) and the
Hong Kong Baptist University Faculty Research Grant (FRG). K. Yang thanks
the support of Institute of Geology and Geophysics, CAS.

\end{document}